**Single Photon Emission from Single Perovskite Nanocrystals of Cesium Lead Bromide**

Fengrui Hu,[1][†] Huichao Zhang,[1,3][†] Chun Sun,[2] Chunyang Yin,[1] Bihu Lv,[1] Chunfeng Zhang,[1] William W. Yu,[2] Xiaoyong Wang,[1][*] Yu Zhang,[2][*]  and Min Xiao[1,4][*]

[1]*National Laboratory of Solid State Microstructures, School of Physics, and Collaborative Innovation Center of Advanced Microstructures, Nanjing University, Nanjing 210093, China*

[2]*State Key Laboratory on Integrated Optoelectronics and College of Electronic Science and Engineering, Jilin University, Changchun 130012, China*

[3]*College of Electronics and Information, Hangzhou Dianzi University, Xiasha Campus, Hangzhou 310018, China*

[4]*Department of Physics, University of Arkansas, Fayetteville, AR 72701, USA*

[†]These authors contributed equally to this work.

[*]Correspondence to X. W. (wxiaoyong@nju.edu.cn), Y. Z. (yuzhang@jlu.edu.cn), and M. X. (mxiao@uark.edu).

**The power conversion efficiency of photovoltaic devices based on semiconductor perovskites has reached ~20% after just several years of research efforts.  With concomitant discoveries of other promising applications in lasers, light-emitting diodes and photodetectors, it is natural to anticipate what further excitements these exotic perovskites could bring about.  Here we report on the observation of single photon emission from single CsPbBr₃ perovskite nanocrystals (NCs) synthesized from a facile**

colloidal approach. Compared with traditional metal-chalcogenide NCs, these $CsPbBr_3$ NCs exhibit nearly two orders of magnitude increase in their absorption cross sections at similar emission colors. Moreover, the radiative lifetime of $CsPbBr_3$ NCs is greatly shortened at both room and cryogenic temperatures to favor an extremely fast output of single photons. The above findings have not only added a novel member to the perovskite family for the integration into current optoelectronic architectures, but also paved the way towards quantum-light applications of single perovskite NCs in various quantum information processing schemes.

Single optical emitters such as natural atoms[1] and molecules[2] are able to emit single photons due to the time interval needed for the electron to recycle between the two-level ground and excited states. This kind of quantum-light emission has been intensively pursued and manipulated since it can provide a robust platform to test fundamental quantum-optical and measurement theories as well as to promote practical quantum-information-processing applications in terms of quantum computing, teleportation and cryptography[3,4]. So far, single photon emission has been discovered in a very limited number of artificially-engineered fluorescent materials ranging from epitaxial quantum dots (QDs)[5], colloidal nanocrystals (NCs)[6], diamond color centers[7] to carbon nanotubes[8], and most recently to $Pr^{3+}$:YAG crystals[9], 4H-SiC wafers[10] and atomically-thin layers of 2D $WSe_2$ (ref. 11). Most of the above single photon sources are associated with structural/compositional defects whose exact origins are sometimes illusive, while only epitaxial QDs and colloidal NCs can be predictably and reproducibly produced with size- and shape-dependent optical properties dictated by the quantum confinement effect. When the selection criterion for single photon sources is raised even higher to be more relevant to practical applications, colloidal NCs are eventually singled out as the unique candidate capable of emitting visible light at room temperature. After nearly two decades of intensive research, a rich spectrum of optical phenomena has been revealed from colloidal NCs at the single particle level[12], the understanding of which is continuously promoting their potential applications such as in biotechnology, medicine, electronics and optoelectronics[13]. One open question of immediate importance is whether a brand new class of colloidal NCs, other than the current ones based almost solely on metal chalcogenides (e.g., CdSe NCs), can be judiciously synthesized to demonstrate its structure- and composition-specific single photon emitting properties.

Semiconductor perovskites can be conveniently described by the formula of ABX$_3$ with A, B and X being an organic or inorganic cation, a metal cation and a halide anion, respectively. Within just the last several years, the power conversion efficiency of solar cells incorporating perovskites as light harvesters has been dramatically increased from the original ~3.8% to the current value of ~20% (refs. 14-20). This stunning performance of perovskites in solar cells benefits largely from their easy solution processing, suitable energy gap and high absorption coefficient, in addition to the high mobility, long diffusion length and slow radiative recombination of photo-excited charge carriers. These superior material and optical properties of perovskites can also be applied to several other optoelectronic devices, with their debut appearances in light emitting diodes[21], lasers[22] and photodetectors[23] having been sequentially demonstrated. To bring the remaining potentials of perovskites into full play, it is imperative to obtain a deeper understanding of the underlying photophysics[24,25] and to fabricate larger-sized single crystals with lower densities of structural defects and carrier traps[26-28]. Meanwhile, with several initial attempts to reduce the size of bulk perovskites to the submicron scale[29,30], colloidal NCs of both cesium and methylammonium lead halides have been recently synthesized[31-33]. These achievements not only provide a flexible tunability for the size-dependent fluorescent colors of perovskites, but also prepare a necessary condition to explore their quantum-light nature of single photon emissions.

Here we focus on optical characterizations of single CsPbBr$_3$ (cesium lead bromide) perovskite NCs whose sizes are close to the exciton Bohr diameter to reach the quantum-confined regime. At room temperature, the photoluminescence (PL) intensity of single CsPbBr$_3$ NCs switched between the "on" and "off" periods intermittently (also called blinking), which can be attributed to random formation of charged excitons and the associated nonradiative Auger recombination. The same Auger process also greatly suppressed the radiative recombination of mult-excitons, leading to almost complete single photon emissions from single CsPbBr$_3$ NCs. At cryogenic temperature, the PL linewidth of a single NC was as broad as ~1 meV under the influence of the spectral diffusion effect. While the above optical behaviors are on a par with those observed in conventional metal-chalcogenide NCs, these CsPbBr$_3$ NCs have exhibited nearly two orders of magnitude increase in their absorption cross sections. Meanwhile, the PL lifetime of single NCs is greatly shortened at both room and cryogenic temperatures to favor an extremely fast output of single photons.

According to a previous report[32], The CsPbBr$_3$ perovskite NCs used in the current experiment were synthesized from a colloidal approach (see Methods) with a cubic shape and an average size of ~9.4 nm as estimated from the transmission electron microscopy measurement (Fig. 1a,b). The NC dimension is close to the Bohr diameter of ~7 nm estimated for the Wannier-Mott excitons of CsPbBr$_3$ perovskites[32] so that the quantum confinement effect can be expected. The first absorption band and the PL peak measured from ensemble CsPbBr$_3$ NCs are located at ~504 nm (~2.46 eV) and ~511 nm (~2.43 eV), respectively (Fig. 1c), corresponding to an energy gap that is blue-shifted by ~200 meV relative to the bulk value of ~2.25 eV (ref. 34).　With pulsed laser excitation at 405 nm, we measured the PL intensity of ensemble CsPbBr$_3$ NCs as a function of the laser power density (Fig. 1d).　For bulk perovskite films, the PL intensity increases almost continuously with the increasing laser power density[35], which is usually accompanied by a blue shift and an obvious broadening of the PL spectrum due to the band-filling effect[24].　Then the PL saturation effect (Fig. 1d) and the invariant PL lineshape (Supplementary Fig. S1) measured for ensemble CsPbBr$_3$ NCs with increasing laser power densities are strong evidences that optical emissions originate from atomic-like, discrete energy levels[7-9].

With the excitation of the same 405 nm pulsed laser, we further performed optical characterizations of single CsPbBr$_3$ NCs at room temperature unless otherwise specified in the text (see Methods).　In Fig. 2a, we plot a confocal scanning PL image acquired from a sample region of 10 μm ×10 μm, where each bright spot corresponds to the optical emission from a single CsPbBr$_3$ NC.　The PL spectra of three single CsPbBr$_3$ NCs are shown in Fig. 2b with an average FWHM (full width at half maximum) of ~100 meV (~25 nm).　Due to the inhomogeneous size distribution (Fig. 1a,b), these PL peaks vary from ~505-515 nm, which is a clear demonstration of the size-dependent energy gaps of the CsPbBr$_3$ NCs.　In Fig. 2c, we also plot the PL intensity of a representative single NC as a function of the laser power density and a similar PL saturation effect to the one shown in Fig. 1d for ensemble NCs can be observed.　With <$N$> representing the average number of photons absorbed per NC per pulse, the PL saturation curve can be fitted with a functional form, $\propto 1 - e^{-<N>} = 1 - e^{-j\sigma}$ (ref. 36), where $\sigma$ and $j$ are the absorption cross section of the NC and the pump fluence of the laser, respectively.　Since the pump fluence $j$ is a controllable parameter in our experiment, the absorption cross section $\sigma$ of this single NC can be reliably estimated to be ~1.23 × 10$^{-13}$ cm$^2$. A similar $\sigma$ of ~2.33 × 10$^{-13}$ cm$^2$ can be obtained from the PL saturation curve shown in Fig. 1d, which can be treated as the average absorption cross section of ensemble $CsPbBr_3$ NCs at the excitation wavelength of 405 nm. Compared to traditional metal-chalcogenide CdSe NCs emitting at similar wavelengths, the $\sigma$ value measured here for $CsPbBr_3$ NCs is enhanced by almost two orders of magnitude (see Methods), indicating that they have inherited the excellent absorption capacity of bulk perovskites.

For convenience, we next performed all the optical characterizations of single $CsPbBr_3$ NCs at $<N> = \sim 0.1$ to avoid any nonlinear effect unless otherwise specified in the text. As measured for a representative $CsPbBr_3$ NC, the PL decay curve can be fitted by a single-exponential function with a lifetime constant of ~6.44 ns (Fig. 2d), which is a typical value for nearly all the single $CsPbBr_3$ NCs studied in our experiment (see Supplementary Fig. S2 for the PL decay curves of two additional NCs). This radiative lifetime of ~6.44 ns is significantly shorter than the counterpart value of hundreds of nanoseconds normally measured for bulk perovskite films[25,26,28], possibly due to the forced overlap of electron and hole wave functions within the quantum-confined volume. In comparison, the PL decay curve measured at room temperature for single metal-chalcogenide CdSe NCs can also be fitted by a single-exponential function but with a radiative lifetime of ~20 ns and beyond[6,37].

The PL time trajectory of a representative $CsPbBr_3$ NC is shown in Fig. 3a with an obvious PL blinking behavior, which was observed from all the single $CsPbBr_3$ NCs studied in our experiment (see Supplementary Fig. S3 for the PL time trajectories of two additional NCs). The probability densities for the time durations of both blinking "on" and "off" periods roughly follow an inverse power-law distribution with an exponent of ~1.5 (Supplementary Fig. S4). Similar to the case of single CdSe NCs[38], the blinking "off" periods can be attributed to nonradiative Auger recombination of charged excitons formed after the photo-excited electron or hole is captured by an external trap and the subsequent excitation of another electron-hole pair in the single NC.   While being neglectable in bulk semiconductors, the Auger effect manifests itself strongly in quantum-confined NCs due to the enhanced Coulomb interaction between charge carriers and the alleviated kinematic restriction on momentum conservation[39].   In fact, the existence of Auger interaction between charge carriers in a single NC is a prerequisite for single photon emission since it provides an efficient channel for nonradiative dissipation of multi-excitons to prevent simultaneous emission of multiple photons[37].   The PL blinking phenomenon recently reported in both $CH_3NH_3PbI_3$ (ref. 40) and $CH_3NH_3PbBr_3$ (ref. 41) perovskite particles, whose dimensions are much larger than the exciton Bohr diameters, might be caused by some other mechanisms instead of Auger recombination of charged excitons since no single photon emission was observed at the single-particle level.

Single photon emissions were observed in all the single $CsPbBr_3$ NCs studied in our experiment from the second-order autocorrelation function $g^{(2)}(\tau)$ measurements.   With the cw excitation from a 400 nm laser, the ratio between $g^{(2)}(0)$ and $g^{(2)}(\tau)$ at long time delays was calculated to be ~0.06 for a representative NC (Fig. 3b).   With the pulsed excitation from the 405 nm laser, the average area ratio between the central $g^{(2)}(0)$ and the side $g^{(2)}(nT)$ peaks was calculated to be ~0.06 for another representative NC (Fig. 3c), where $n \geq 1$ is an integer and $T$ is the laser repetition time.   These excellent photon anti-bunching features strongly confirm that single $CsPbBr_3$ NCs can serve perfectly as a single photon source.   In our experiment, we also increased the laser power density to set $<N>$ larger than ~1.0 so that the possibility for the generation of multi-excitons in a single NC should be greatly increased[37,39].   As can be seen in Supplementary Fig. S5 for the pulsed excitation of a representative NC, there was almost no increase in the average area ratio between the central and the side peaks with the increasing laser power density. This implies that Auger recombination of multi-excitons is extremely efficient in single $CsPbBr_3$ NCs, thus rendering a pure single photon stream without random interruptions of multi-photon events.

We have also performed preliminary optical characterizations of $CsPbBr_3$ NCs at the cryogenic temperature of ~4 K. Consistent with what was reported previously for bulk $CH_3NH_3PbI_3$ perovskites[35], the PL peak measured at ~4 K for ensemble $CsPbBr_3$ NCs showed no blue shift when compared to the one measured at room temperature (Supplementary Fig. S6). In Fig. 4a, we plot the PL spectra of four single $CsPbBr_3$ NCs with the calculated FWHMs that are all slightly smaller than 1 meV. These relatively broad PL spectra reflect the fact that spectral diffusion is still drastic at ~4 K, as can be seen from the time-dependent PL spectral image shown in Fig. 4b for two single $CsPbBr_3$ NCs. The PL decay curve of a representative $CsPbBr_3$ NC is shown in Fig. 4c, which can be fitted well only by a bi-exponential function with a fast and a slow lifetime component of ~355 ps and ~5.75 ns, respectively (see Supplementary Fig. S7 for the PL decay curves of two additional $CsPbBr_3$ NCs). For single CdSe NCs, the PL decay also becomes bi-exponential at cryogenic temperatures due to electronically-coupled optical emissions from a bright state and a lower-lying dark state with a radiative lifetime of hundreds of nanoseconds[42]. The energy-level splitting between the above two states can reach the scale of several meV so that the long-lived dark state cannot be easily depopulated by the thermal energy at cryogenic temperatures, leading to its dominant contribution to the total optical emissions over the bright state in CdSe NCs. If thermal mixing of bright and dark states can really be invoked to explain the PL dynamics observed here for single $CsPbBr_3$ NCs, their energy-level splitting must be significantly smaller than that in CdSe NCs since the slow lifetime component of ~5.75 ns accounts for at most ~5% of the total PL decay.

Based on all the experimental results obtained so far, we would like to point out several advantages of the newly-synthesized perovskite NCs over traditional metal-chalcogenide NCs. First, to achieve the same brightness, the laser power density used to excite a single perovskite NC can be decreased by almost two orders of magnitude due to the enhanced absorption cross section. This would greatly reduce the possibility of tissue damage in biolabeling and bioimaging studies, the scattered laser light under resonant or near-resonant excitation conditions, and the working current in electrically-driven lightening devices. Second, although the perovskite NCs studied here have a PL peak around ~510 nm, the sizes and compositions can be further manipulated to render broad fluorescent colors spanning the green-to-red wavelength range of traditional metal-chalcogenide NCs. Moreover, the PL peaks of perovskite NCs can be potentially extended to even shorter wavelengths, e.g., with the $CsPbCl_3$ (ref. 32) and $CH_3NH_3PbCl_3$ (ref. 33) compositions, for the realizations of blue light-emitting diodes and single photon sources. Third, the fast radiative lifetimes at both room and cryogenic temperatures, together with the suppressed multi-photon emission at high laser power densities, guarantee that single perovskite NCs can be used as a highly-efficient, pure single photon source driven at high repetition rates of ~1 GHz.

To summarize, we have synthesized colloidal $CsPbBr_3$ NCs and successfully demonstrated their capability of emitting single photons. The introduction of these novel optical emitters to the colloidal NC and semiconductor perovskite families will surely stimulate intensive research efforts in future works of material synthesis, fundamental science and device application. Since the perovskite NCs come into play in the research community just very recently, there is much room for the optimization of the material properties by borrowing nearly two decades of synthesis protocols developed for metal-chalcogenide NCs. For example, proper selection of the passivation ligands and wise adoption of core/shell structures would certainly help to boost the fluorescent quantum efficiency with robust photo-stability, as well as to suppress or even eliminate the PL blinking and spectral diffusion effects. From the fundamental science of view, a lot of intriguing issues need to be addressed for single perovskite NCs especially at cryogenic temperatures, including the fine-structure splitting of bright and dark states, as well as the magneto-optical and the coherent properties. Given the fact that the fabrication techniques for solar cells, light-emitting diodes and photodetectors based on bulk perovskite films have been fully developed, it would also be interesting to see how perovskite NCs behave in these optoelectronic devices.

**Methods**

**Synthesis of colloidal CsPbBr$_3$ NCs**

*Chemicals：* Cs$_2$CO$_3$ (99.9%) was purchased from J&K. Oleic acid (90%, OA) and octadecene (90%, ODE) were purchased from Alfa Aesar. PbBr$_2$ (99.0%) and oleylamine (80-90%, OLA) were purchased from Aladdin. Toluene (99.5%) was purchased from Beijing Chemical Factory.

*Preparation of Cs-oleate:* Cs$_2$CO$_3$ (0.8 g), OA (2.5 mL) and ODE (30 mL) were loaded into a 100 mL 3-neck flask and dried under vacuum for 1 h at 120 ℃. The reaction solution was then kept at 150 ℃ under N$_2$ until it became clear.

*Synthesis and purification of CsPbBr$_3$ NCs:* ODE (10 mL) and PbBr$_2$ (0.138g) were added into a 50 mL 3-neck flask and dried under vacuum for 1 h at 120 ℃. Then, dried OLA (1 mL) and dried OA (1 mL) were injected under N$_2$. After the mixture became clear, the temperature was raised to 180 ℃ and the Cs-oleate solution (0.8 mL, 0.1 M in ODE, preheated to 100 ℃) was quickly injected. The reaction mixture was kept at this temperature for 5 s, and then was cooled down to room temperature by an ice-water bath. The final NCs were purified by centrifugation and re-dispersed in toluene to form a long-term colloidally-stable solution. The solution absorption and emission measurements were taken with a V-550 UV-Visible spectrophotometer from Jasco, and the fluorescent quantum yield was estimated to be ~58%.

## Optical characterizations of $CsPbBr_3$ NCs

One drop of the concentrated or diluted solution of $CsPbBr_3$ NCs was spin-casted onto a fused silica substrate to form a solid film for the optical characterizations of ensemble or single NCs at room temperature. The 405 nm output of a 5 MHz, picosecond diode laser or the 400 nm output of a cw laser was used as the excitation source. The laser beam was focused onto the sample substrate by an immersion-oil objective (N.A.=1.4). PL signal of the ensemble NCs or a single NC was collected by the same objective and sent through a 0.5 m spectrometer to a charge-coupled-device camera for the PL spectral measurements. The PL signal of a single NC can be alternatively sent through a non-polarizing 50/50 beamsplitter to two avalanche photo diodes (APDs) in a time-correlated single photon counting (TCSPC) system with a time resolution of ~250 ps. The TCSPC system was operated under the time-tagged, time-resolved mode so that the arrival times of each photon relative to the laboratory time and the laser pulse time could be both obtained, which allowed us to plot the PL time trajectory and the PL decay curve, respectively. Moreover, the delay times between photons collected by one APD and those by the other could be summed up to yield the second-order autocorrelation $g^{(2)}(\tau)$ functions.

For the low-temperature optical characterizations of ensemble or single $CsPbBr_3$ NCs, the sample substrate was contained in a helium-free cryostat operated at ~4 K. Very similar optical setups to those described above were employed except that the immersion-oil objective was replaced by a dry objective (N.A.=0.8) and the 485 nm output of a 5.6 MHz, picosecond supercontinuum fiber laser was used as the excitation source.

**Calculations of absorption cross sections**

The PL saturation curve of either ensemble or single CsPbBr$_3$ NCs can be fitted with a functional form, $\propto 1 - e^{-<N>} = 1 - e^{-j\sigma}$, where $\sigma$ and $j$ are the absorption cross section and the laser pump fluence, respectively. The pump fluence $j$ can be calculated from $j = P/FE$, where the laser power density $P$ can be directly measured, $F$ is the laser repetition rate of 5 MHz, and $E$ is the laser photon energy at 405 nm. Based on the above procedures, the absorption cross sections estimated for ensemble CsPbBr$_3$ NCs and a representative CsPbBr$_3$ NC are ~2.33 × 10$^{-13}$ cm$^2$ and ~1.23 × 10$^{-13}$ cm$^2$, respectively. For comparison, commercial Qdot525 colloidal NCs from Invitrogen have a peak emission wavelength of ~525 nm and a molar extinction coefficient $\varepsilon$ of 360000 cm$^{-1}$M$^{-1}$ at 405 nm. The absorption cross section $\sigma$ of Qdot525 NCs can be calculated to be ~1.38 × 10$^{-15}$ cm$^2$ from $\sigma = \ln(10)\frac{1000}{N_A}\varepsilon$, where $N_A$ is the Avogadro's number.

**Acknowledgements**

This work is supported by the National Basic Research Program of China (Nos. 2012CB921801 and 2011CBA00205), the National Natural Science Foundation of China (Nos. 91021013, 91321105, 11274161 and 11321063), Jiangsu Provincial Funds for Distinguished Young Scientists (No. BK20130012), and the PAPD of Jiangsu Higher Education Institutions.

**Author contributions**

X.W., C.Z., and M.X. conceived and designed the experiments.  C.S., W.Y. and Y.Z. prepared the samples.  F.H., H.Z., C.Y. and B.L. performed the optical experiments.  F.H., Y.Z. and X.W. analyzed the data.  X.W., Y.Z. and M.X. co-wrote the manuscript.

**Additional information**

The authors declare no competing financial interests. Correspondence and requests for materials should be addressed to X.W., Y.Z. or M.X.

**Figure Legends**

**Figure 1 | Structural and optical properties of ensemble CsPbBr$_3$ NCs.   a,** Transmission electron microscopy of CsPbBr$_3$ NCs.   **b,** Histogram for the size distribution of CsPbBr$_3$ NCs. **c,** Solution absorption and emission spectra measured for ensemble CsPbBr$_3$ NCs.   **d,** PL intensity of ensemble CsPbBr$_3$ NCs measured as a function of the laser power density $P$ and fitted with the function, $\propto 1 - e^{-\alpha P}$, where $\alpha$ is a fitting constant related to the absorption cross section.

**Figure 2 | Basic PL properties of single CsPbBr$_3$ NCs at room temperature.   a,** Confocal scanning PL image of single CsPbBr$_3$ NCs. The scanning step was 100 nm and the dwelling time at each pixel is 100 ms.   **b,** PL spectra of three single CsPbBr$_3$ NCs obtained with an integration time of 10 s.   **c,** PL intensity of a single CsPbBr$_3$ NC measured as a function of the laser power density $P$ and fitted with the function, $\propto 1 - e^{-\alpha P}$, where $\alpha$ is a fitting constant related to the absorption cross section.   **d,** PL decay curve of a single CsPbBr$_3$ NC fitted by a single-exponential function with a radiative lifetime of ~6.44 ns.

**Figure 3 | PL blinking and single photon emission of single CsPbBr$_3$ NCs at room temperature.   a,** PL time trajectory of a single CsPbBr$_3$ NC with a binning time of 100 ms. **b,** Second-order autocorrelation function $g^{(2)}(\tau)$ measurement of a single CsPbBr$_3$ NC under the cw excitation.   **c,** Second-order autocorrelation function $g^{(2)}(\tau)$ measurement of a single CsPbBr$_3$ NC under the pulsed excitation.

**Figure 4 | Basic PL properties of single CsPbBr$_3$ NCs at ~4 K.   a,** PL spectra of four single CsPbBr$_3$ NCs obtained with an integration time of 10 s.   The associated FWHM of each PL spectrum is also shown.   **b,** Time-dependent PL spectral image for two single CsPbBr$_3$ NCs

where the PL blinking and spectral diffusion effects can be clearly observed. The integration time for each PL data point is 10 s.   **c,** PL decay curve of a single CsPbBr$_3$ NC fitted by a bi-exponential function with a fast and a slow lifetime of ~355 ps (~98.4%) and ~5.75 ns (~1.6%), respectively.

Figure 1:

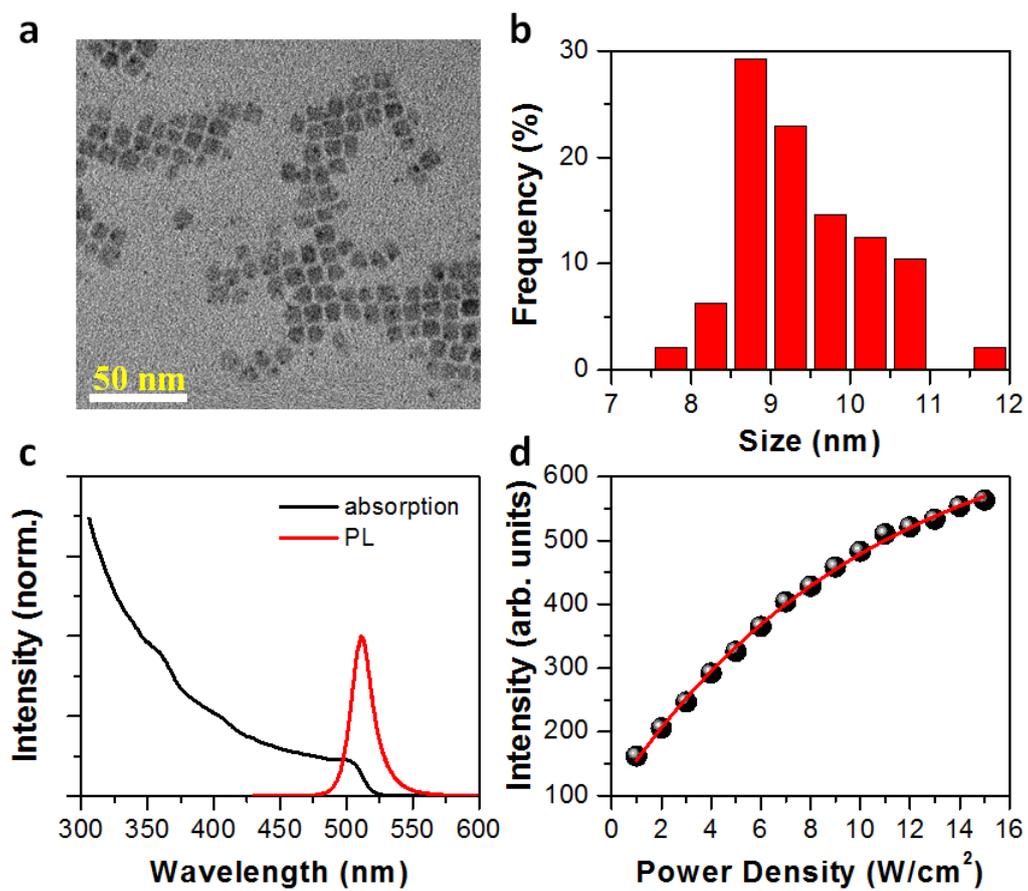

Figure 2:

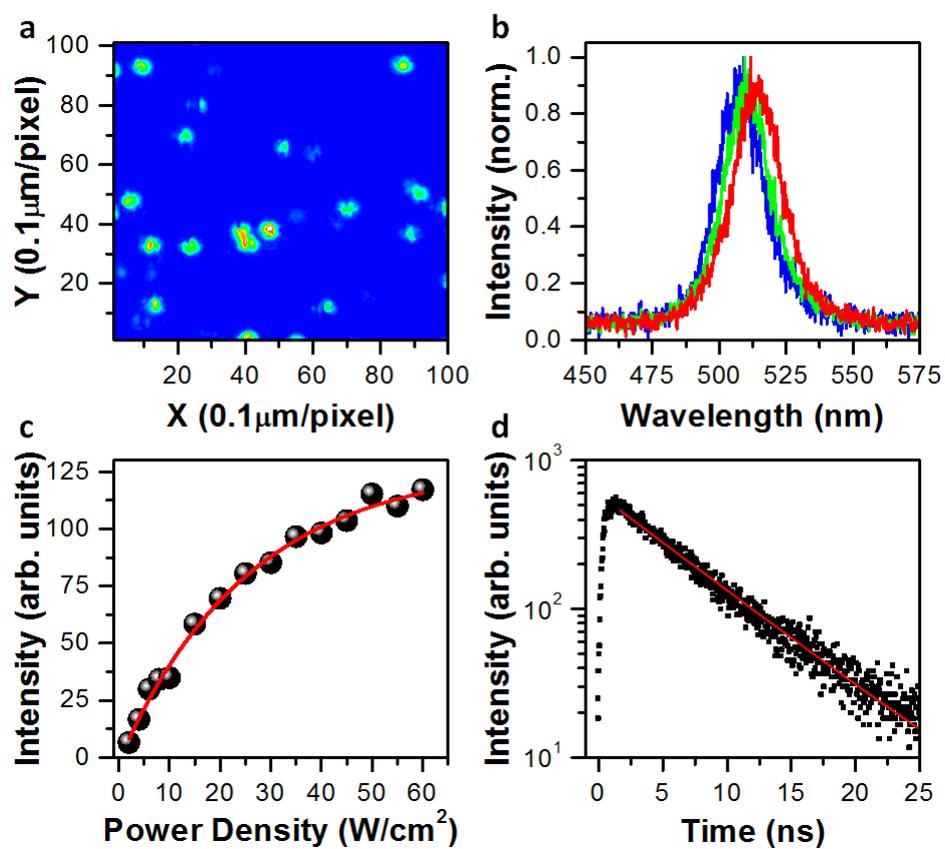

Figure 3

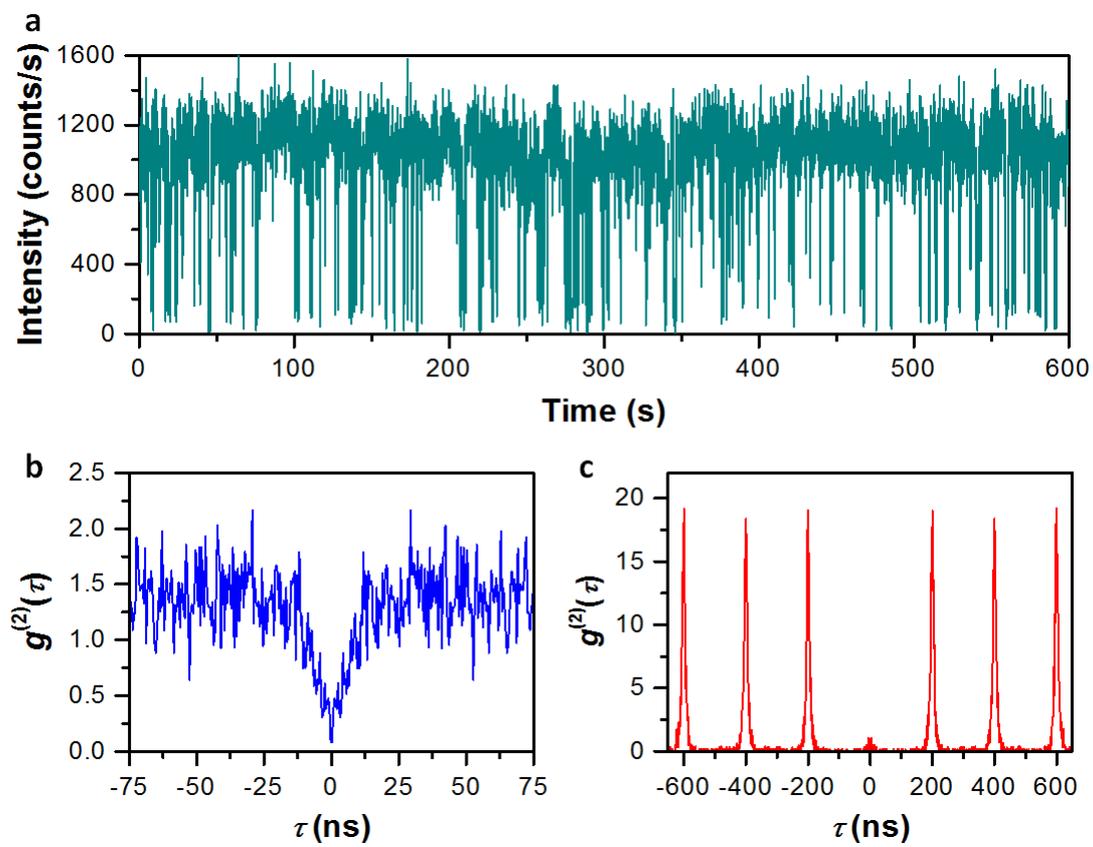

Figure 4:

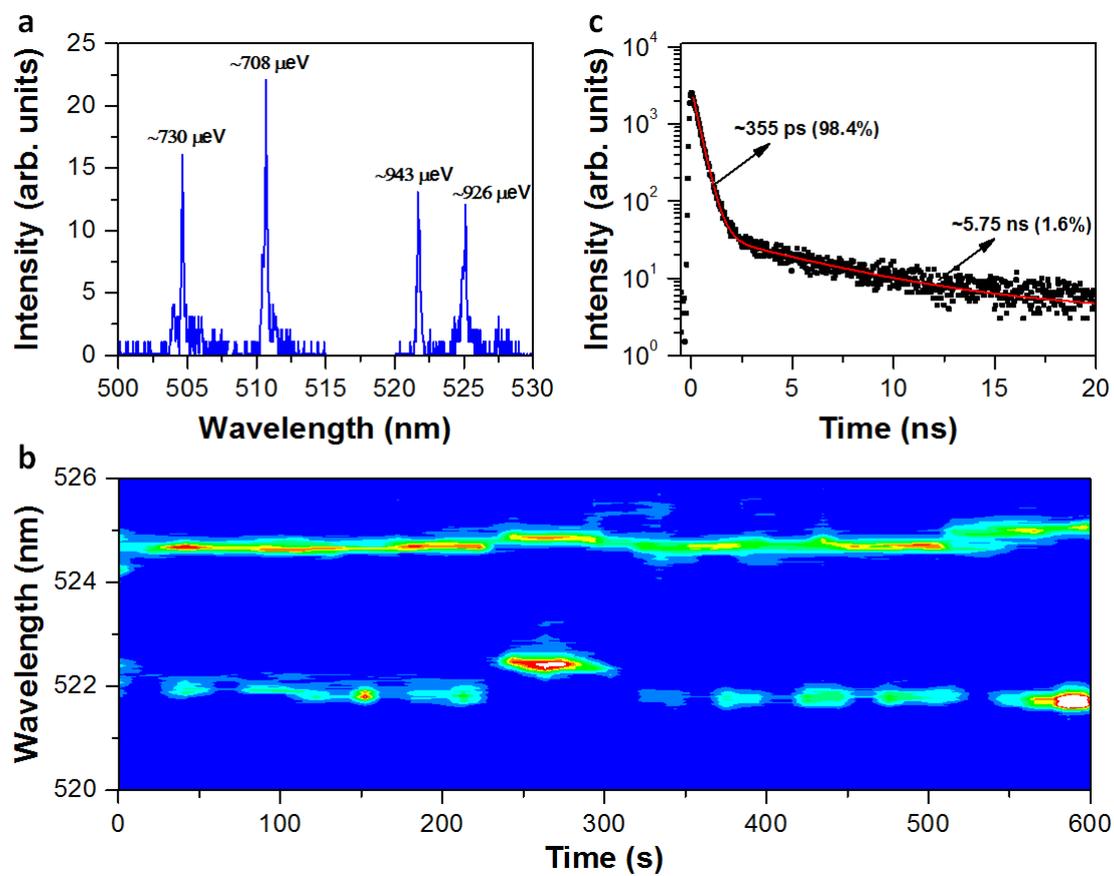

# Single Photon Emission from Single Perovskite Nanocrystals of Cesium Lead Bromide

Fengrui Hu,[1†] Huichao Zhang,[1,3†] Chun Sun,[2] Chunyang Yin,[1] Bihu Lv,[1] Chunfeng Zhang,[1] William W. Yu,[2] Xiaoyong Wang,[1*] Yu Zhang,[2*] and Min Xiao[1,4*]

[1]*National Laboratory of Solid State Microstructures, School of Physics, and Collaborative Innovation Center of Advanced Microstructures, Nanjing University, Nanjing 210093, China*

[2]*State Key Laboratory on Integrated Optoelectronics and College of Electronic Science and Engineering, Jilin University, Changchun 130012, China*

[3]*College of Electronics and Information, Hangzhou Dianzi University, Xiasha Campus, Hangzhou 310018, China*

[4]*Department of Physics, University of Arkansas, Fayetteville, AR 72701, USA*

*These authors contributed equally to this work.

[*]Correspondence to X. W. (wxiaoyong@nju.edu.cn), Y. Z. (yuzhang@jlu.edu.cn), and M. X. (mxiao@uark.edu).

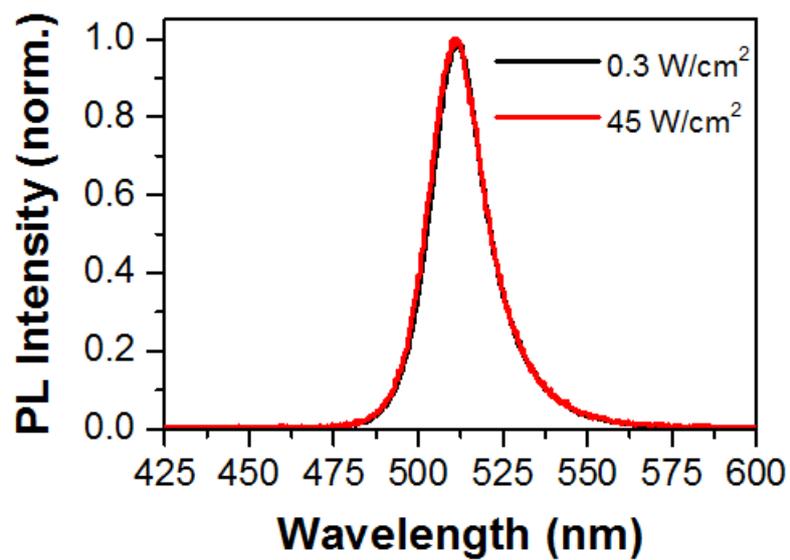

**Figure S1.** PL spectra of ensemble CsPbBr₃ NCs measured at the laser power densities of 0.3 W/cm² and 45 W/cm², respectively.

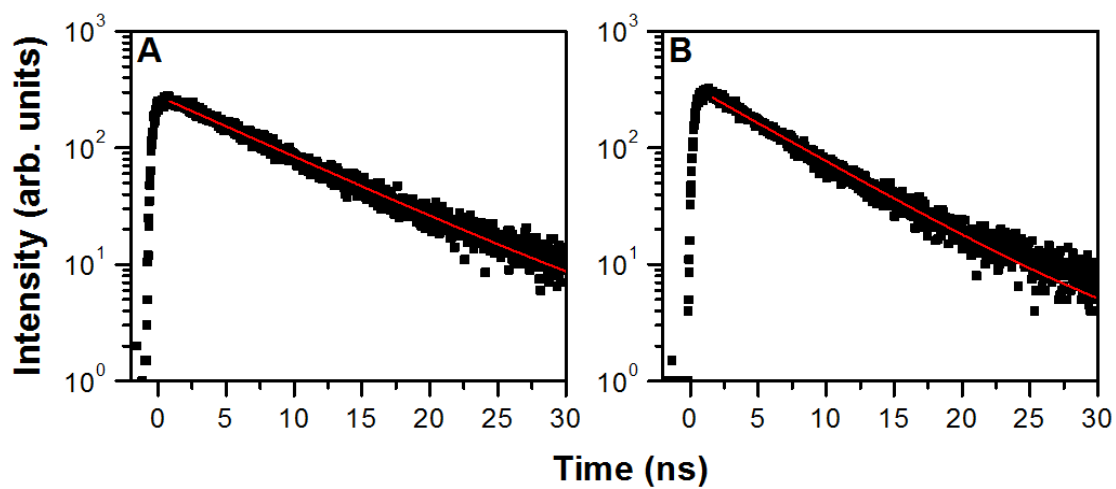

**Figure S2.** PL decay curves of two single CsPbBr$_3$ NCs fitted by single exponential functions with radiative lifetimes of **(A)** ~8.28 ns and **(B)** ~6.57 ns, respectively.

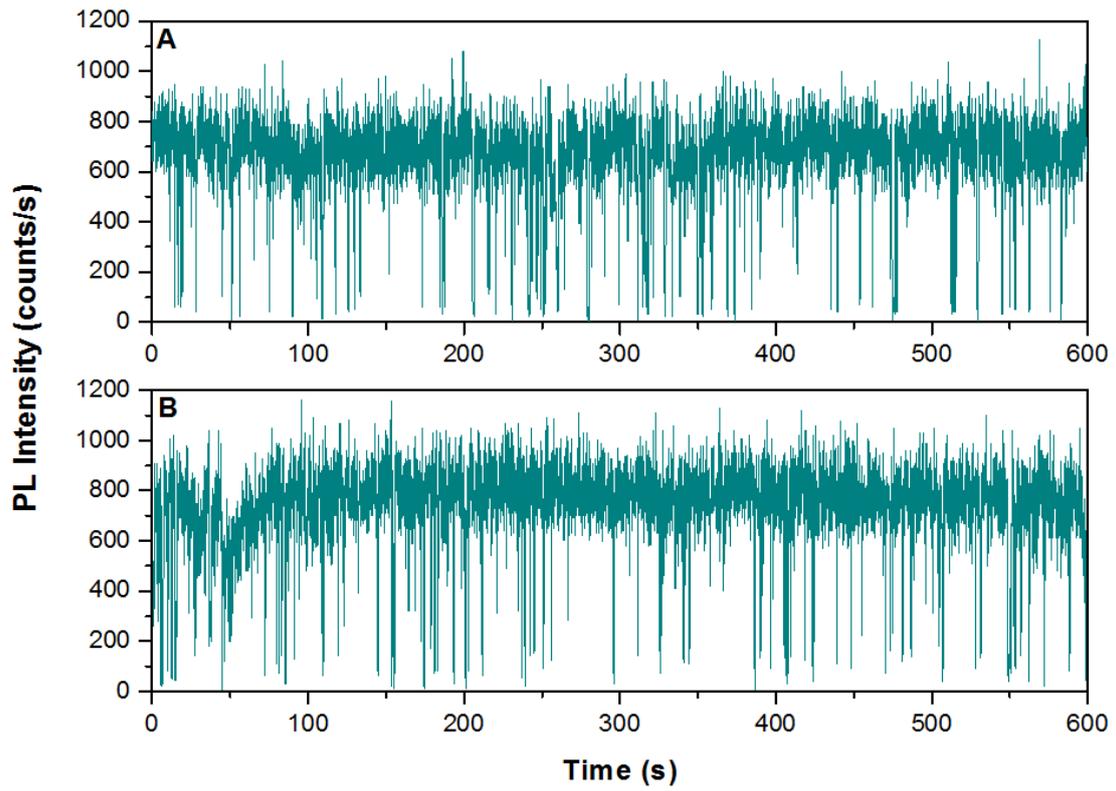

**Figure S3.** PL time trajectories of two single CsPbBr$_3$ NCs with a binning time of 100 ms. The two single NCs in both (**A**) and (**B**) were excited at a laser power density corresponding to <$N$> = ~0.1.

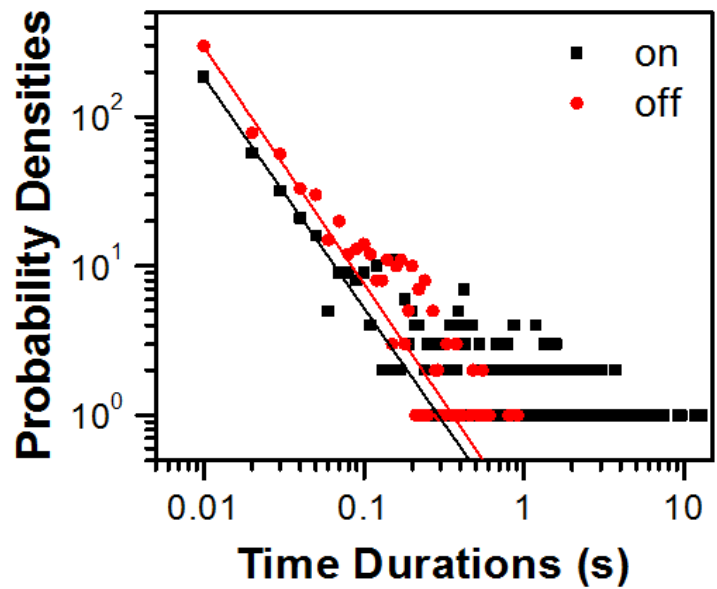

**Figure S4.** Probability densities for the distributions of the blinking "on" and "off" time durations of a represetative CsPbBr$_3$ NC, which can both be roughly fitted with inverse power-law functions with the exponents of ~1.59 and ~1.55, respectively.

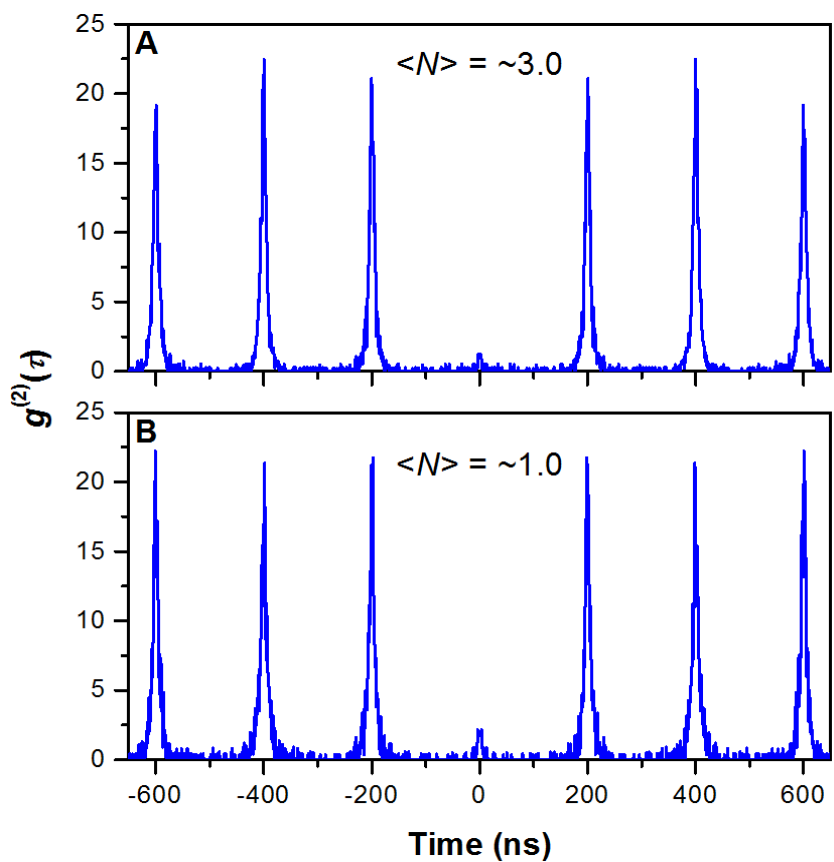

**Figure S5.** Second-order autocorrelation function $g^{(2)}(\tau)$ measurements of a representative CsPbBr$_3$ NC excited with the pulsed laser power densities corresponding to (**A**) $<N> = \sim3.0$ and (**B**) $<N> = \sim1.0$, with the average area ratios between the central $g^{(2)}(0)$ and the side $g^{(2)}(nT)$ peaks being $\sim0.09$ and $\sim0.05$, respectively.

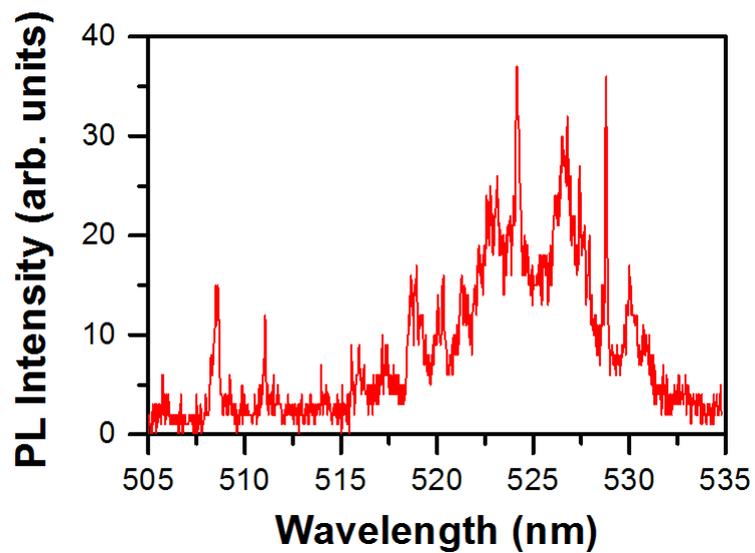

**Figure S6.** PL spectrum measured at ~4 K for a low-density sample of ensemble CsPbBr₃ NCs, where the narrow PL lines from single NCs can be clearly resolved.

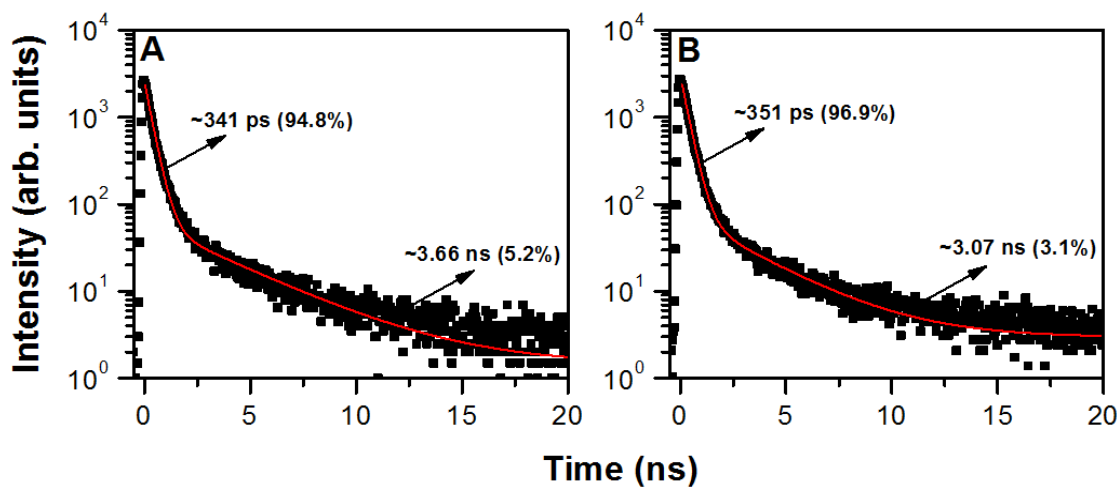

**Figure S7.** PL decay curves of two single CsPbBr$_3$ NCs fitted both by biexponential functions. The fast and slow lifetimes are ~341 ps (~94.8%) and ~3.66 ns (~5.2%) for the single NC in **(A)**, and they are ~351 ps (~96.9%) and ~3.07 ns (~3.1%) for the single NC in **(B)**.